\pdfoutput=1
\documentclass[a4paper,11pt]{article}

\usepackage{jheppub}
\usepackage[T1]{fontenc}
\usepackage{bbold}
\usepackage{xcolor}
\usepackage{tikz}
\usepackage{amsfonts}
\def\beq{\begin{equation}}
\def\eeq{\end{equation}}
\def\beqn{\begin{eqnarray}}
\def\eeqn{\end{eqnarray}}
\def\bpm{\begin{pmatrix}}
\def\epm{\end{pmatrix}}

\title{Gravitational Entropy}

\author[a]{Sangmin Choi}
\affiliation[a]{Institute for Theoretical Physics, University of Amsterdam, PO Box 94485, 1090 GL Amsterdam, The Netherlands.}
\author[b,c,d]{and Malcolm J.\ Perry}
\affiliation[b]{School of Physics and Astronomy, Queen Mary University of London, Mile End Road, London E1 4NS, UK.}
\affiliation[c]{DAMTP, Centre for Mathematical Sciences, Wilberforce Road, Cambridge, CB3 0WA, UK.}
\affiliation[d]{Trinity College, Cambridge, CB2 1TQ, UK.}
\emailAdd{s.choi@uva.nl}
\emailAdd{malcolm@maths.cam.ac.uk}

\abstract{
	We formulate the classical gravitational entropy of a horizon as a Noether charge that does not require the notion of a temperature,
		and which is applicable to horizons that are not necessarily associated with black holes.
	This introduces a correction to the covariant phase space formalism that accounts for the configuration-dependence of the generating vector field conjugate to the charge.
	The vector field is related to the proposal of Bousso that the gravitational entropy of a region is determined by the lightsheet at its boundary.
	We test the formula on various black hole and cosmological horizons.
}

\begin{document}

\maketitle
\flushbottom

\section{Introduction}
\label{sec:intro}

Understanding black holes as objects that obey the laws of thermodynamics has a long history, going back over 50 years \cite{Bardeen:1973gs,Bekenstein:1973ur,Hawking:1975vcx}.
The first law of thermodynamics of a rotating, charged black hole states that the variations of its horizon area $A$, mass $M$, angular momentum $J$, and charge $Q$ are related by
\begin{align}
	\delta M
	&=
		\frac{\kappa}{8\pi G} \delta A
		+ \Omega_H \delta J
		+ \Phi \delta Q
\end{align}
where $\kappa$ is the surface gravity, $\Omega_H$ is the angular velocity and $\Phi$ is the electric potential of the black hole.
Comparing to the thermodynamics of a rotating, charged body, the term involving the area behaves like the heat transfer $T\delta S$ at temperature $T$, where $\delta S$ is the change in
	the entropy:
	the surface gravity and the area of the horizon correspond to the Hawking temperature and the Bekenstein-Hawking entropy of the black hole, respectively.

For a black hole whose event horizon is a bifurcate Killing horizon,
	Wald's Noether charge method can be used to derive the first law from the diffeomorphism charge associated with a Killing vector field $\xi$ \cite{Wald:1993nt,Jacobson:1993vj,Iyer:1994ys}.
Generalizations to dynamical scenarios have been proposed in \cite{Dong:2013qoa,Wall:2015raa,Mishra:2017sqs,Hollands:2024vbe,Gao:2001ut}.
For perturbations around stationary black holes, the quantity $T\delta S$ can be written as an integral of the Noether charge 2-form associated with $\xi$ over the bifurcation surface.
This carries a factor of the surface gravity $\kappa$, and so this procedure is tied to the notion of the Hawking temperature.

In this paper, we relax the temperature dependence and derive the entropy $\delta S$ from the Noether charge, making it applicable to cases that do not involve black holes.
At first glance, one may argue that this is achieved by simply including an overall factor $1/T$ in the vector field whose conjugate diffeomorphism charge leads to $T\delta S$.
However, as we shall see below, the presence of this factor restricts perturbations only to nearby black hole solutions with identical surface gravity, i.e.\ $\delta\kappa=0$.
We take the configuration space to be the space of the same type of black hole solutions, with varying values of parameters such as $M$, $J$ and $Q$.
Then, the surface gravity (and hence the Hawking temperature) depends on the point in the configuration space, and therefore in general $\delta \kappa\neq0$.
To derive the entropy $\delta S$ as a Noether charge conjugate to a vector field $\xi$, one has to take into account the possibility for $\delta\xi\neq 0$.
This leads to a new term appearing in the covariant phase space method, which has been discussed independently in \cite{Ashtekar:2024stm}.
The phase space dependence of the vector field has been considered earlier in the literature; see for instance \cite{Gao:2003ys,Frodden:2017qwh,Adami:2021nnf,Chandrasekaran:2021vyu}.
It has also appeared indirectly in the form of modified Lie brackets in \cite{Barnich:2001jy,Barnich:2011mi}.

The main result of this paper is the application of this framework to horizons in stationary spacetimes, where the entropy $\delta S$ (and thus $S$) is derived from the Noether charge conjugate to
	the null generator of the horizon normalized in a universal manner.
This is in accordance with Bousso's observation \cite{Bousso:2002ju} that entropy should be determined by the behavior of the generators of the lightsheets on the boundary of the region of concern.
As a consistency check, we demonstrate that the formula yields the correct entropy for Kerr and Kerr-Newman black holes.
Since our formula does not involve the notion of Hawking temperature, it is applicable to spacetimes that do not involve the notion of a black hole.
For black holes with a bifurcation Killing horizon and a well-defined notion of mass and angular momentum at infinity, the integrated entropy $S$ is given simply as an integral of
	the Noether charge 2-form on any section of the horizon.
Interestingly, applying the same construction to cosmological horizons, this integral yields the correct Gibbons-Hawking entropy \cite{Gibbons:1977mu} for de Sitter and Kottler spacetimes,
	which is useful since de Sitter spacetime lacks parameters with respect to which we can vary the metric.

The paper is organized as follows.
In section \ref{sec:CPS}, we review and extend the covariant phase space formalism and diffeomorphism charges to account for vector fields that are functions on the configuration space.
In section \ref{sec:GC}, we apply the formalism to Einstein gravity.
We illustrate the universal normalization of the null vector field in the Schwarzschild black hole in section \ref{sec:Schw}, and apply it to the Kerr black hole in section \ref{sec:Kerr}.
In section \ref{sec:EM} we briefly discuss adding electromagnetic sources, and then in section \ref{sec:KN} we work out the entropy of Kerr-Newman black holes.
We work through example spacetimes that exhibit cosmological horizons in sections \ref{sec:DeSitter} and \ref{sec:SdS}.
We end with a discussion of the results in section \ref{sec:Conc}.
Some details have been delegated to the appendix.

\section{Covariant Phase Space} \label{sec:CPS}

The starting point of our discussion is the action principle for a set of fields $\phi$. $\phi$ 
will always
include the gravitational field described by the metric tensor $g_{ab}$, but may also include 
other fields, such as the electromagnetic field $A_a$ that we will consider in later sections. 
The action $I$ is the integral
of a four-form Lagrangian $L$ so 
\begin{align}
	I = \int L(\phi).
\end{align}
The action has dimensions of $[M][L]$ so that, when inserted into the path integral, $I/\hbar$ is 
dimensionless.

Variation of the fields $\phi\rightarrow\phi+\delta\phi$ induces a variation of the action $\delta I$
where
\begin{align}
	\delta I =  \int E(\phi)\cdot\delta\phi + d\theta(\phi,\delta\phi).
\end{align}
The equation of motion is $E(\phi)=0$, but in general, there will also be a boundary term
that defines $\theta(\phi,\delta\phi)$,  the presymplectic potential three-form.

Gravitational theories are invariant under infinitesimal diffeomorphisms generated by a vector field $\xi^a$.
The resultant transformations on the various components of $\phi$ are given by  
their Lie derivative with respect to $\xi$, so that in general $\delta\phi = {\cal L}_\xi\phi$.

One can find a formula for the Noether charge conjugate to $\xi$. Starting from the 
presymplectic potential $\theta$, we make a second variation of $\phi$ given by $\delta^\prime\phi$ so that  the presymplectic form $\omega$ is
\begin{align}
	\omega(\phi,\delta\phi,\delta^\prime\phi) = \delta\theta(\phi,\delta^\prime\phi)
	 - \delta^\prime\theta(\phi,\delta\phi)
\end{align}
where $(\delta\delta'-\delta'\delta)\phi=0$.
Now let the variation $\delta^\prime={\cal L}_\xi$ be a diffeomorphism.
Provided $E(\phi)=0$ and $\delta\phi$ obeys the linearised equations of motion,  $\omega$ is closed
and can be written as $\omega=d\hat G$. Then 
\begin{align}
	\delta Q_\xi =\int_{\Sigma_s} \omega(\phi,\delta\phi,{\cal L}_\xi\phi)= \int_S \hat G, \label{eq:deltaQ}
\end{align}
where $Q_\xi$ is the Noether charge conjugate to $\xi$, $\Sigma_s$ is a spacelike three-surface, and $S=\partial\Sigma_s$ is its two-dimensional boundary.
The expression $\delta Q_\xi$ should be interpreted as the change in the charge conjugate to $\xi$ as 
the fields $\phi$ vary into $\phi + \delta\phi$.

The Noether current resulting from the diffeomorphism generated by $\xi$ is the three-form $\hat J[\xi]$ defined by
\begin{align}
	\hat J[\xi] = \theta(\phi,{\cal L}_\xi \phi) - \iota_\xi L(\phi). \label{eq:noethercurrent}
\end{align}
$\hat J$ is closed provided $\phi$ obeys the equation of motion $E(\phi)=0$. One can then write
\begin{align}
	\hat J = d\hat F
\end{align}
for some two-form $\hat F[\xi]$ that is a functional of both $\phi$ and $\xi$.
The variation of the current is then
\begin{align}
	\delta\hat J[\xi] = \delta\theta(\phi,\delta\phi)-\iota_\xi d\theta(\phi,\delta\phi) -\iota_{\delta\xi} L(\phi), \label{eq:nother} 
\end{align}
provided $E(\phi)=0$ holds.
The last term on the (\ref{eq:nother}) accounts for the possibility that $\xi^a$ is not constant in the configuration space, in which case $\delta\xi^a$ is not identically zero.
Cartan's magic identity for an arbitrary $p$-form $X$ is
\begin{align}
	{\cal L}_\xi X = d\iota_\xi X + \iota_\xi d X,
\end{align}
and so we find that
\begin{align}
	\delta\hat J[\xi] = \delta\theta(\phi,{\cal L}_\xi\phi)-{\cal L}_\xi\theta(\phi,\delta\phi)+d\iota_\xi \theta(\phi,\delta\phi) -\iota_{\delta\xi} L(\phi). \label{eq:varnother}
\end{align}
Since $\delta\xi$ is not necessarily zero, the first two terms on the right-hand side of (\ref{eq:varnother})  are related to the presymplectic three-form $\omega$ plus a correction that is linear in ${\cal L}_{\delta\xi}\phi$,
\begin{align}
	\delta\theta(\phi,{\cal L}_\xi\phi) - {\cal L}_\xi\theta(\phi,\delta\phi) = \omega(\phi,\delta\phi,{\cal L}_\xi\phi) + \theta(\phi,{\cal L}_{\delta\xi}\phi).
\end{align}
Thus the variation of the Noether current becomes
\begin{align}
	\delta\hat J[\xi] = \omega(\phi,\delta\phi,{\cal L}_\xi\phi) + \theta(\phi,{\cal L}_{\delta\xi}\phi)+d\iota_\xi \theta(\phi,\delta\phi) -\iota_{\delta\xi} L(\phi), 
\end{align}
which can be reorganized into the form
\begin{align}
	\omega(\phi,\delta\phi,{\cal L}_\xi\phi) = \delta\hat J[\xi] -d\iota_\xi \theta(\phi,\delta\phi) - \hat J[\delta\xi]. \label{eq:psw}
\end{align}
Now consider a spacelike three-surface $\Sigma_s$ with boundary $S$.
The variation of the Noether charge is
\begin{align}
	\delta Q_\xi = \int_{\Sigma_s} \omega(\phi,\delta\phi,{\cal L}_\xi \phi) = \int_{\Sigma_s} \delta\hat J[\xi] - d\iota_\xi \theta(\phi,\delta\phi)-\hat J[\delta\xi]. 
\end{align}
Provided $E(\phi)=0$ holds, we can write $\hat J=d\hat F$ and so by Stokes' theorem
\begin{align}
	\delta Q_\xi = \int_S \delta\hat F[\xi] - \iota_\xi \theta(\phi,\delta\phi)-\hat F[\delta\xi]. \label{deltaQ}
\end{align}
The last term $\hat F[\delta\xi]$ in \eqref{deltaQ} accounts for the phase space dependence of the vector field; see Appendix \ref{appendix}.
 
Notice that there is a consistency condition here that needs to be satisfied.
For the expression on the r.h.s.\ to be consistent, we require that the last two terms be a total variation. That is,
\begin{align}
	\int_S \iota_\xi\theta(\phi,\delta\phi)+\hat F[\delta\xi] = \delta \int_S C \label{eq:consistency}
\end{align}
where $C$ is some two-form (which may not be covariant) that is determined, up to the addition of $dW$ for some one-form $W$.
If such a $C$ exists, then the integrated entropy $Q_\xi$ exists and takes the simple form
\begin{align}
	Q_\xi = \int_S (\hat F[\xi] - C)
	.
\end{align}
If it is the case that no such $C$ exists, then $Q_\xi$ does not exist. 

It seems as if the restrictions resulting from this consistency condition are rather stringent. However, 
the definitions of both the action and the presymplectic potential are fraught with ambiguity \cite{Wald:1999wa,Iyer:1994ys}. 
The first ambiguity lies in the observation that the action is not unique.
The equations of motion are invariant under a change  of the action
$L \rightarrow L + dZ$ for some three-form $Z$. A second ambiguity is that the Noether current can be 
modified simply by the addition 
of the exterior derivative of some two-form $Y$ so that $\hat J \rightarrow \hat J +dY$.
This would appear to make  $Q_\xi$ arbitrary. 
One might suppose that there is another ambiguity in the definitions of $\hat F$ and $\hat  G$ in that one could add  to them
pieces that are the exterior derivatives of some one-forms. But provided that $S$ is closed, 
this will affect neither $Q_\xi$  
nor $\delta Q_\xi$ so we will not pursue that possibility. 
Summarizing these two induced transformations on the various differential forms we have encountered
so far, we note that
\begin{align}
	&L \rightarrow L+dZ \\
	&\theta \rightarrow \theta + \delta Z + dY \\
	&\hat J \rightarrow \hat J + dY + d(\iota_\xi Z) \\
	&\hat F \rightarrow \hat F + Y + \iota_\xi Z \\
	&\hat G \rightarrow \hat G + \delta Y(\phi,{\cal L}_\xi\phi) - {\cal L}_\xi Y(\phi,\delta\phi).
\end{align}
The effect of these transformations on the charge and its variation can easily be determined and we find that
\begin{align}
	\delta Q_\xi  \rightarrow \delta Q_\xi + \int_S \left(\delta Y(\phi,{\cal L}_\xi\phi)-{\cal L}_\xi Y(\phi,\delta\phi)\right).
\end{align}
We observe that, while one may have a non-zero $Z$ for reasons outlined by Gibbons, Hawking and York \cite{York:1972sj,Gibbons:1976ue},
	such a term does not enter into the expression for the charge.
$Y$ can be chosen in a more or less arbitrary fashion, and the existence of the charge $Q_\xi$ depends on a consistency condition analogous to \eqref{eq:consistency}.
We expect that these issues will not give rise to complications in our analysis, since we restrict our attention to exact symmetries in stationary spacetimes \cite{Wald:1999wa}.

\section{Gravitational Charges}
\label{sec:GC}

The Einstein-Hilbert action $I_\text{EH}$ is the usual starting point for establishing the equations of motion
 in general relativity.
\begin{align}
	I_\text{EH} = \frac{1}{16\pi G} \int_{\cal M} \ (R-2\Lambda)\  \sqrt{-g}\  d^4x
\end{align}
where the integral is taken over the spacetime manifold ${\cal M}$. If one make a variation of the metric 
$g_{ab} \rightarrow g_{ab}+h_{ab}$, then we obtain the both the equations of motion
\begin{align}
	R_{ab}=\Lambda g_{ab}
\end{align}
and the presymplectic potential 
\begin{align}
	\theta^a_\text{(basic)} = \frac{1}{16\pi G} ( \nabla_b h^{ab} - \nabla^a h)
	\label{eq:theta_grav}
\end{align}
where $h=h_{ab}g^{ab}$.
Given the equations of motion, we can find the linearised equations of motion for $h_{ab}$
\begin{align}
		\Box h_{ab}
		+ \nabla_a\nabla_b h
		- \nabla_b\nabla_c h^c_a
		- \nabla_a\nabla_c h^c_b
		+ 2R_{acbd}h^{cd}
	&=
		0
	.
\end{align}

We are interested here in surfaces that are spacelike so that we can measure the charge
contained in a closed two-surface surrounding some region of space.  The Noether current coming from the 
basic part of the action is 
\begin{align}
	\hat J^a[\xi] = \frac{1}{16\pi G}\biggl(\Box\,\xi^a - \nabla^a\nabla_b\,\xi^b + R^{ab}\xi_b - \xi^aR +2\Lambda\xi^a\biggr)
	.
\end{align}
Provided the background equations of motion $R_{ab}=\Lambda g_{ab}$ holds, $\nabla_a\hat J^a =0$ and so $\hat J^a = \nabla_b\hat F^{ab}$.
We choose
\begin{align}
	\hat F^{ab}[\xi] = \frac{1}{16\pi G} ( \nabla^b\xi^a - \nabla^a\xi^b)
	.
\end{align}
The charge conjugate $Q_\xi$ then satisfies
\begin{align}
	\delta Q_\xi = \frac12\int_S \ \Bigl( \delta \hat F^{ab}[\xi]-2\theta^a\xi^b-\hat F^{ab}[\delta\xi]\Bigr)  dS_{ab}.
\end{align}
The presymplectic form is
\begin{multline}
	\omega^a(h,h^\prime)
	=
		\frac{1}{16\pi G} \Biggl[
			\tfrac{1}{2}h^\prime\nabla^ah
			- \tfrac{1}{2}h\nabla^ah^\prime
			- \tfrac{1}{2}h^\prime\nabla_bh^{ab}
			+ \tfrac{1}{2}h\nabla_b h^\prime{}^{ab}
			- \tfrac{1}{2}h^\prime{}^{ab}\nabla_bh
			\\
			+ \tfrac{1}{2}h^{ab}\nabla_bh^\prime
			- \tfrac{1}{2}h^\prime{}^{bc}\nabla^a h_{bc}  
			+ \tfrac{1}{2}h^{bc}\nabla^a h^\prime_{bc}
			+ h^\prime_{bc}\nabla^bh^{ac}
			- h_{bc}\nabla^bh^\prime{}^{ac}
		\Biggr].
\end{multline}
Putting $h^\prime_{ab}={\cal L}_\xi g_{ab} = \nabla_a\xi_b + \nabla_b\xi_a$, using the equations of motion and the 
linearised equations of motion for $h_{ab}$, we find that $\omega^a=\nabla_b\hat G^{ab}$ where
\begin{align}
	\hat G^{ab}
	\equiv
		\frac{1}{16\pi G}\bigg[
			\xi^b\nabla^a h
			&- \xi^a\nabla^b h
			+ \xi^a\nabla_c h^{bc}
			- \xi^b\nabla_c h^{ac}
			- \xi_c\nabla^ah^{bc}
			\nonumber\\&
			+ \xi_c\nabla^bh^{ac}
			- \frac12h\nabla^a \xi^b
			+ \frac12h\nabla^b \xi^a
			- h^{bc}\nabla_c\xi^a
			+ h^{ac}\nabla_c\xi^b
		\bigg]
	.
\end{align}
From this expression we see that the variation $\delta Q_\xi$  of the charge $Q_\xi$
as the metric varies from $g_{ab}$ to $g_{ab}+h_{ab}$ is
\begin{align}
	\delta Q_\xi = \frac{1}{2} \int \hat G^{ab}dS_{ab}
	,
	\label{eq:dQfromhG}
\end{align}
as one would expect.

\section{The Schwarzschild Case}
\label{sec:Schw}

Our aim in this section is to develop a candidate expression for the gravitational entropy.
We will do this by an examination of the geometry of the Schwarzschild black hole
and conjecture a general result for the appropriate Noether charge. In subsequent sections, 
we will test our conjecture.  

The Schwarzschild metric in $(t,r,\theta,\phi)$ coordinates takes the familiar form
\begin{align}
	ds^2 = -V(r)dt^2 + {dr^2\over V(r)} + r^2(d\theta^2 +  \sin^2\theta d\phi^2)
\end{align}
where
\begin{align}
	V(r) = 1 - {2GM\over r}.
\end{align}
A section of the future horizon is the obvious location of a two-surface to see if one can find a 
$\xi$ that reproduces the known black hole entropy. 
Thus our surface  $\Sigma_s$  will stretch from spacelike infinity and intersect the future horizon at some 
moment of advanced time. To explore this scenario, 
we introduce ingoing Eddington-Finkelstein coordinates $(v,r,\theta,\phi)$ with the advanced time being $v$
given by $v = t + r^\ast$ and $dr^\ast/dr = 1/V(r)$.
The metric then takes the form
\begin{align}
	ds^2 = -V(r)dv^2 +2dvdr + r^2(d\theta^2 + \sin^2\theta d\phi^2).
\end{align}
$v$ is a null coordinate that labels time on the future horizon.
The surface $\Sigma_s$ intersects the horizon at some $v=v_0$ and $r=2GM.$
The spacetime is static and so ${\partial/\partial v}= k^a {\partial/\partial x^a}$
where $k^a$ is a Killing vector that is null and geodesic on the horizon. 
The surface gravity on the horizon $\kappa$ is defined by
\begin{align}
	k^a \nabla_a k^b = \kappa k^b
\end{align}
indicating the $k^a$ is not affinely parametrised by $v$. Evaluating $\kappa$ yields 
\begin{align}
	 \kappa = \frac{1}{4GM}.
\end{align}
The Hawking temperature $T_H$
for black holes is universally given by
\begin{align}
	T_H = \frac{\hbar\kappa}{2\pi}
\end{align}
and so for Schwarzschild
\begin{align}
	T_H = \frac{\hbar}{8\pi GM}.
\end{align}

We now need to find the correct vector $\xi^a$ to give the black hole entropy. 
Recall that the charge as defined in previous sections has the dimension of $[M]$, so to find a
dimensionless entropy, we need to rescale $Q_\xi$ by a factor of $1/\hbar$. The entropy
would be given by
\begin{align}
	S= \frac{l_p}{\hbar}Q_\xi.
\end{align}
The choice of $\xi$ at first seems to be completely undetermined.

Bousso suggested some time ago \cite{Bousso:2002ju} that entropy should be determined
by the behavior of the generators of lightsheets on the boundary of the region of concern. 
Accordingly, we pick $\xi^a$ to be the null generator of the horizon.
Thus,  $\xi^a$ is some multiple of $k^a$, so we set
$\xi^a=\lambda k^a$, where $\lambda$ may depend on $M$.
Then $\delta\xi^a = (\delta\lambda) k^a$.
Now we determine $Q_\xi$.
For this, we first compute
\begin{align}
	\hat F_{vr}[\xi]= -\frac{M\lambda }{8\pi r^2},\qquad \hat F_{vr}[\delta\xi]= -\frac{M \delta\lambda}{8\pi r^2},
\end{align}
and
\begin{align}
	\theta_v=-\frac{\delta M}{8\pi r^2},\qquad \theta_r=0,\qquad \theta_\theta=0,\qquad \theta_\phi=0
	.
\end{align}
On the horizon in $(v,r,\theta,\phi)$ coordinates we can choose the unit timelike vector $t^a=({1/\sqrt{V}},0,0,0)$ and the unit spacelike vector $s^a=(1/\sqrt{V},\sqrt{V},0,0)$ so that $s^at_a=0$.
Integrating on the horizon and recalling
\begin{align}
	dS^{ab} = -2\,t^{[a}\,s^{b]} \ \sigma^{\tfrac{1}{2}} \ d^2x
	,
\end{align}
so that
$dS^{vr}=-r^2\sin\theta d\theta d\phi$,
we find the total charge obeys
\begin{align}
	\delta Q_\xi
	=
		\frac12\left[
			\delta (\lambda M)
			+ \lambda\delta M
			- M\delta\lambda
		\right]
	= \lambda\delta M.
\end{align}
The choice $\lambda = 2\pi(\kappa l_p)^{-1}=8\pi GMl_p^{-1}$ then gives $\delta Q_\xi = \delta(4\pi G M^2l_p^{-1})$ and
\begin{align}
	S = \frac{l_p}{\hbar}Q_\xi =  \frac{4\pi GM^2}{\hbar} = \frac{A}{4\hbar G}
\end{align}
where $A$ is the area of the intersection of $\Sigma_s$ with the horizon.
With this choice the consistency condition \eqref{eq:consistency} is met with $C=0$.

We are therefore led to conjecture that in general $\xi^a$ should be chosen as the null geodesic generator of the lightsheet of the region whose
gravitational entropy we wish to find. It needs to obey the null geodesic
equation and its parametrisation is fixed by
\begin{align}
	\xi^a\nabla_a \xi^b = \frac{2\pi}{l_p} \xi^b
	.
	\label{eq:parameterisation_xi}
\end{align}
We adopt this normalization throughout the paper.
This should be understood as a conjectured prescription inspired by Bousso's lightsheet construction \cite{Bousso:2002ju}, rather than a result derived from first principles.

Now look at the variation $\delta Q_\xi $. A variation in the mass  of the
black hole $\delta M$ results in an $h_{ab}$  whose only non-vanishing
component is $h_{vv} = {2G\delta M/r}$. A computation of $\hat G_{vr}$ results
in
\begin{align}
	\hat G_{vr} = -\frac{2G M\delta M}{l_pr^2}.
\end{align}
then
\begin{align}
	\delta Q_\xi = \frac{8\pi G}{l_p}M\delta M
\end{align}
so
\begin{align}
	\delta S = \frac{8\pi GM\delta M}{\hbar}.
\end{align}
This result is consistent with the first law of black hole thermodynamics 
and is also consistent with the evaluation of $Q_\xi$.

In the following sections, we apply this formula to compute the entropy of Kerr and Kerr-Newman black holes.
These examples should be regarded as nontrivial consistency checks of the proposal, rather than independent derivations of black hole entropy.

\section{The Kerr Case}
\label{sec:Kerr}

In this section, we test our entropy formula for the Kerr black hole.
The metric of a Kerr black hole of mass $M$ and angular momentum $J$ in Boyer-Lindquist coordinates takes the form
\begin{align}
	ds^2
	&=
		-\frac\Delta\Sigma(dt-a\sin^2\theta d\phi)^2
		+ \frac\Sigma\Delta dr^2
		+ \Sigma d\theta^2
		+ \frac{\sin^2\theta}\Sigma\left(
			adt - (r^2+a^2)d\phi
		\right)^2
\end{align}
where $r_s=2GM$ is the Schwarzschild radius, $a=J/M$ is the angular momentum per mass, $\Sigma=r^2+a^2\cos^2\theta$ and $\Delta=r^2-r_sr+a^2$.
We change to a set of coordinates $(v,r,\theta,\phi)$ that is the Kerr analog of the ingoing Eddington-Finkelstein coordinates,
\begin{align}
	dv = dt + \frac{(r^2+a^2)}\Delta dr
	,\qquad
	d\phi^\text{new} = d\phi^\text{old} + \frac a\Delta dr
	,
	\label{eq:KerrEF}
\end{align}
where $v$ is a null coordinate that labels time on the horizons.
In these coordinates, the Kerr metric can be organized into the following form
\begin{align}
	ds^2
	&=
		-\left(1-\frac{r_s r}\Sigma\right)(dv - a\sin^2\theta d\phi)^2
		+ 2(dv - a\sin^2\theta d\phi)(dr - a\sin^2\theta d\phi)
		\nonumber\\&\quad
		+ \Sigma (d\theta^2+\sin^2\theta d\phi^2)
	.
\end{align}
The volume element is $\sqrt{-g}=\Sigma\sin\theta$.
The only non-zero components of the inverse metric are
\begin{align}
	g^{vv}
	&=
		\frac{a^2}{\Sigma} \sin^2\theta
	,\qquad
	g^{vr}
	=
		\frac{r^2+a^2}{\Sigma}
	,\qquad
	g^{rr}
	=
		\frac\Delta\Sigma
	,\\
	g^{v\phi}
	\nonumber &=
	g^{r\phi}
	=
		\frac a\Sigma
	,\qquad
	g^{\theta\theta}
	=
		\frac1\Sigma
	,\qquad
	g^{\phi\phi}
	=
		\frac1{\Sigma\sin^2\theta}
	.
	\label{eq:Kerrig}
\end{align}
The radii of the inner and outer horizons $r_-$ and $r_+$ are located at the solutions to $\Delta=0$,
\begin{align}
	r_\pm = \frac12(r_s\pm \sqrt{r_s^2-4a^2})
	.
\end{align}
The surface $\Sigma_s$ intersects the outer horizon $r=r_+$ at some $v=v_0$.

The Killing vector $k^a\partial_a = \partial_v + \Omega_+ \partial_\phi$ is null and geodesic on the outer horizon satisfying $\xi^a\nabla_a\xi^b = \kappa \xi^b$,
where
\begin{align}
	\Omega_+ = \frac{a}{r_+^2+a^2}
\end{align}
is the angular speed on the outer horizon.
The surface gravity $\kappa$ has the expression
\begin{align}
	\kappa = \frac{r_+-r_-}{2r_sr_+}.
\end{align}
The Hawking temperature is given by $T_H = \hbar\kappa/2\pi$.

To find the entropy, we define $\xi^a$ to be the vector obtained by rescaling $k^a$ according to the normalization \eqref{eq:parameterisation_xi}.
It has the components
\begin{align}
	\xi^a\partial_a = \frac{2\pi}{l_p\kappa}\left(\frac{\partial}{\partial v} + \Omega_+ \frac{\partial}{\partial \phi}\right)
	.
	\label{eq:Kerr_xi}
\end{align}
Let us first check that the consistency condition \eqref{eq:consistency} holds.
One finds that the integral of $-\iota_\xi\theta$ takes the simple form
\begin{align}
	\int_S -\iota_\xi\theta
	&=
		\frac{\pi}{l_p\kappa}\delta M
	\label{eq:Kerr_c1}
	.
\end{align}
Since the vector field $\xi^a$ \eqref{eq:Kerr_xi} depends on $M$ and $J$, its variation $\delta\xi^a$ is non-vanishing,
\begin{align}
	\delta\xi^a\partial_a  = \delta\Big(\frac{2\pi}{l_p\kappa}\Big)\frac{\partial}{\partial v} + \delta\Big(\frac{2\pi\Omega_+}{l_p\kappa}\Big)\frac{\partial}{\partial \phi}
	.
\end{align}
This implies that the charge conjugate to $\xi$ obtains a contribution of the form
\begin{align}
	\int_S -\hat F[\delta\xi]
	&=
		- \frac{M}{2}\delta\Big(\frac{2\pi}{l_p\kappa}\Big)
		+ J \delta\Big(\frac{2\pi\Omega_+}{l_p\kappa}\Big)
	\label{eq:Kerr_c2}
	.
\end{align}
The consistency condition \eqref{eq:consistency} requires that the sum of \eqref{eq:Kerr_c1} and \eqref{eq:Kerr_c2} be a total variation.
It turns out that the sum vanishes:
\begin{align}
	\int_S -\iota_\xi\theta-\hat F[\delta\xi]
	=
		\frac{\pi}{l_p\kappa}\delta M
		- \frac{M}{2}\delta\Big(\frac{2\pi}{l_p\kappa}\Big)
		+ J \delta\Big(\frac{2\pi\Omega_+}{l_p\kappa}\Big)
	= 0,
	\label{eq:Kerr_c3}
\end{align}
which can be derived by a straightforward computation, using the variations
\begin{align}
	\delta\Omega_+
	&=
		-\frac{2G a}{r_s^2r_+}\frac{(r_++\frac12 r_s)}{(r_+-\frac{1}{2}r_s)}\delta M
		+ \frac{2G}{r_s}\frac{1}{(r_+^2-a^2)} \delta J
	,\\
	\delta \Big(\frac{1}{\kappa}\Big)
	&=
		- \frac{\delta \kappa}{\kappa^2}
	=
		\frac{8 G a r_s}{(r_+-r_-)^3}
		\left[
			\delta J
			+ \frac{r_+}{2ar_s}
			\left(
				4r_+^2-3r_s^2
			\right)\delta M
		\right]
	.
\end{align}
This is a consequence of the first law of black hole thermodynamics
\begin{align}
	\delta M =T_H\delta S+ \Omega_+ \delta J,
\end{align}
and the Smarr formula \cite{Smarr:1972kt}, which for the Kerr black hole takes the form
\begin{align}
	M=2T_HS + 2\Omega_+ J
	.
\end{align}
Taking a variation of the Smarr formula and using the first law, one can show that the following identity holds,
\begin{align}
	\frac{\delta M}{2T_H}-\frac{M}{2}\delta\Big(\frac{1}{T_H}\Big)+J\delta\Big(\frac{\Omega_+}{T_H}\Big)=0,
\end{align}
which is, after putting $T_H=\hbar \kappa/2\pi$, equivalent to equation \eqref{eq:Kerr_c3}.
Thus, the consistency condition \eqref{eq:consistency} is met with $C=0$ just as was the case for Schwarzschild.

Since the contributions from $\iota_\xi\theta$ and $\hat F[\delta\xi]$ to the Noether charge collectively vanish,
\begin{align}
	\delta Q_\xi
	&=
		\int_S \delta \hat F[\xi] - \iota_\xi\theta -\hat F[\delta\xi]
	=
		\delta \int_S \hat F[\xi]
	,
\end{align}
the full charge given just by the integral of $\hat F[\xi]$, which we find to have the expression
\begin{align}
	Q_\xi=\int_S \hat F[\xi]=\frac{2\pi}{l_p\kappa}\left(\frac{M}{2}-\Omega_+J\right).
\end{align}
After putting $T_H = \hbar\kappa/2\pi$, we obtain the entropy
\begin{align}
	S
	&=
		\frac{l_p}{\hbar}Q_\xi
	=
		\frac{M}{2T_H}
		- \frac{\Omega_+J}{T_H}
	=
		\frac{A}{4G\hbar}
	,
\end{align}
where $A=4\pi(r_+^2+a^2)$ is the area of the outer Kerr horizon.

Now let's look at the variation of the Noether charge.
The variations $\delta M$ and $\delta J$ to the mass and angular momentum of Kerr black hole result in the following non-vanishing components of $h_{ab}$:
\begin{align}
	h_{vv}
	&=
		\frac{2G r}{\Sigma^2}
		\left[
			r^2\delta M
			+ (3a\delta M - 2\delta J) a\cos^2\theta
		\right]
	,\\
	h_{v\phi}
	&=
		- \frac{2G r}{\Sigma^2}
		\sin^2\theta
		\left[
			r^2\delta J
			+ (2a\delta M - \delta J) a^2\cos^2\theta
		\right]
	,\\
	h_{r\phi}
	&=
		\frac{\sin^2\theta}{M}(a\delta M-\delta J)
	,\\
	h_{\theta\theta}
	&=
		-\frac{2a}{M}\cos^2\theta(a\delta M-\delta J)
	,\\
	h_{\phi\phi}
	&=
		-\frac{2a^2}{M}\sin^2\theta
		\left[
			1
			+ \frac{GMr}{\Sigma^2}\sin^2\theta(r^2-a^2\cos^2\theta)
		\right]
		\delta M
		\nonumber\\&\quad
		+ \frac{2a}{M}\sin^2\theta
		\left(
			1+ \frac{2GM r^3}{\Sigma^2}\sin^2\theta
		\right) \delta J
	.
\end{align}
Using these expressions, a computation of the integral of $\hat G$ on $S$ yields the charge
\begin{align}
	\delta Q_\xi
	&=
		\int_S \hat G
	=
		\frac{\hbar}{l_pT_H}\left(
			\delta M - \Omega_+ \delta J
		\right)
	.
\end{align}
This is consistent with the variation $\delta S=\frac{l_p}{\hbar}\delta Q_\xi$ one obtains from the first law of black hole thermodynamics.

\section{Electromagnetism and Kerr-Newman black holes}

\subsection{Including Electromagnetism}\label{sec:EM}

In electromagnetism, the basic field from which everything else is built is the vector potential $A_a$.
It gives rise to a field strength tensor $F_{ab}=\nabla_aA_b-\nabla_bA_a$ which is invariant under the 
gauge transformation $A_a \rightarrow A_a+\partial_a\epsilon$ for arbitrary $\epsilon.$ The electromagnetic
action is
\begin{align}
	I_\text{(em)} = -\frac{1}{4} \int F_{ab}F^{ab}\ \sqrt{-g}\ d^4x.
\end{align}
Routine calculations yield the energy-momentum tensor $T_{ab}$ as
\begin{align}
	T_{ab} = F_a{}^cF_{bc} - \tfrac{1}{4}g_{ab}F_{cd}F^{cd} \label{eq:EMT}
\end{align}
together with the Maxwell equations
\begin{align}
	\nabla_bF^{ab} = 0. \label{eq:emeom}
\end{align}
The Maxwell equations need to be supplemented by the Bianchi identity $\nabla_{[a}F_{bc]}=0$, which is a direct consequence of the definition of $F_{ab}$ in terms of $A_a$.

Following the prescription outlined for gravity, we find an extra contribution to the presymplectic potential 
that must be added to the  gravitational contribution \eqref{eq:theta_grav}
\begin{align}
	\theta^a_\text{(em)} = - F^{ab} \delta A_b .
\end{align}
The electromagnetic field then gives rise to some extra contributions to the various quantities we have discussed
in the purely gravitational case (see also \cite{Gao:2001ut,Hajian:2015xlp}).
There will therefore be extra terms in the gravitational Noether current
which can be derived exactly as in the purely gravitational case. 
Also, in addition to diffeomorphism, the theory has a $U(1)$ gauge symmetry.
Taking this into account, we find the Noether current conjugate to the diffeomorphism $\xi$ and $U(1)$ gauge transformation $\epsilon$ to be
\begin{align}
	\hat J^a_\text{(em+grav)}[\xi,\epsilon]
	&=
		\hat J^a_\text{(grav)}[\xi]
		- F^{ab}({\cal L}_\xi A_b - \partial_b \epsilon)
	.
\end{align}
As before, if the Einstein equations and the Maxwell equations are both satisfied, $\hat J^a_\text{(em+grav)}$ is conserved
and yields 
\begin{align}
	\hat F^{ab}_\text{(em+grav)} = \hat F^{ab}_\text{(grav)} - (\xi^c A_c + \epsilon) F^{ab}.
\end{align}
It is straightforward to show that the presymplectic three-form is related to the current conjugate to $\xi$ and $\epsilon$ in a way analogous to the gravitational case \eqref{eq:psw},
\begin{align}
	\omega(\phi, \delta\phi, {\cal L}_\xi\phi + \delta_\epsilon\phi)
	&=
		\delta\hat J[\xi,\epsilon]
		- d\iota_\xi \theta(\phi,\delta\phi)
		- \hat J[\delta \xi,\delta\epsilon]
	,
\end{align}
where $\delta_\epsilon$ denotes the $U(1)$ gauge transformation $\delta_\epsilon A_a = \partial_a \epsilon$, $\delta_\epsilon g_{ab}=\delta_\epsilon F_{ab}=0$.
Likewise, the Noether charge conjugate to $\xi$ and $\epsilon$ can be computed using $\hat F[\xi,\epsilon]$,
\begin{align}
	\delta Q_{\xi,\epsilon}
	&=
		\int_S
		\delta \hat F[\xi,\epsilon]
		- \iota_\xi\theta(\phi,\delta\phi)
		- \hat F[\delta\xi,\delta\epsilon]
	.
	\label{eq:dQwCharge}
\end{align}
The presymplectic form is
\begin{align}
	\omega_\text{(em+grav)}^a
	&=
		\omega_\text{(grav)}^a
		+ \frac12h' F^{ab}\delta A_b
		+ h'^{ac}F^b{}_c \delta A_b
		- h'^{bc}F^a{}_c \delta A_b
		+ g^{ac}g^{bd}\delta' F_{cd} \delta A_b
		\nonumber\\&\quad
		- \frac12h F^{ab}\delta' A_b
		- h^{ac}F^b{}_c \delta' A_b
		+ h^{bc}F^a{}_c \delta' A_b
		- g^{ac}g^{bd}\delta F_{cd} \delta' A_b
	.
\end{align}
Putting $h'_{ab} = {\cal L}_\xi g_{ab}$, $\delta' A_a = {\cal L}_\xi A_a + \partial_a \epsilon$ and using the equations of motion as well as the linearised equations of motion for $h_{ab}$ and $\delta A_a$,
	we find that $\omega_\text{(em+grav)}^a = \nabla_b\hat G^{ab}_\text{(em+grav)}$ where
\begin{align}
	\hat G^{ab}_\text{(em+grav)}
	&\equiv
		\hat G^{ab}_\text{(em)}
		- \xi^a F^{bc} \delta A_c
		+ \xi^b F^{ac} \delta A_c
		- \xi^c F^{ab} \delta A_c
		\nonumber\\&\quad
		- \left(
			\frac12h F^{ab}
			+ h^{ac}F^b{}_c
			- h^{bc}F^a{}_c
			+ g^{ac}g^{bd}\delta F_{cd}
		\right)(\xi^eA_e+\epsilon)
	.
	\label{eq:hGwithCharge}
\end{align}

\subsection{Kerr-Newman Black Hole}\label{sec:KN}

Now we apply our formula to the Kerr-Newman black hole.

In the Boyer-Lindquist coordinates, the Kerr-Newman metric of mass $M$, angular momentum $J$ and electric charge $Q$ reads
\begin{align}
	ds^2
	&=
		- \frac\Delta\Sigma(dt-a\sin^2\theta d\phi)^2
		+ \frac\Sigma\Delta dr^2
		+ \Sigma d\theta^2
		+ \frac{\sin^2\theta}{\Sigma}
		\left[
			a dt
			- (r^2+a^2)d\phi
		\right]^2
	\label{eq:KNmetric0}
\end{align}
where $r_s=2GM$ is the Schwarzschild radius, $a=J/M$ is the angular momentum per mass, $\Sigma$ and $\Delta$ are defined as
\begin{align}
	\Sigma = r^2 + a^2\cos^2\theta
	,\qquad
	\Delta = r^2-r_sr + a^2+r_Q^2
	.
	\label{eq:KNSigmaDelta}
\end{align}
Here $r_Q$ is the length scale corresponding to the electric charge, defined as
\begin{align}
	r_Q^2 = \frac{G}{4\pi}Q^2,
\end{align}
in units with unit vacuum permittivity.
Just like in the Kerr case, we perform a change of coordinates \eqref{eq:KerrEF} but instead with the new definition of $\Delta$ for Kerr-Newman given in \eqref{eq:KNSigmaDelta}.
This takes us to the Kerr-Newman analog $(v,r,\theta,\phi)$ of the ingoing Eddington-Finkelstein coordinates, in terms of which the metric \eqref{eq:KNmetric0} reads
\begin{align}
	ds^2
	&=
		- \left(
			1
			- \frac{(r_sr-r_Q^2)}{\Sigma}
		\right) dv^2
		+ 2dvdr
		- \frac{2a}{\Sigma}(r_sr-r_Q^2)\sin^2\theta dvd\phi
		- 2a\sin^2\theta drd\phi
		+ \Sigma d\theta^2
		\nonumber\\&\quad
		+ \frac{\sin^2\theta}{\Sigma}
		\left[
			(r^2+a^2)^2
			-\Delta a^2\sin^2\theta
		\right] d\phi^2
	.
\end{align}
The inverse metric takes the same form as those \eqref{eq:Kerrig} for the Kerr metric except for the difference in the definition of $\Delta$ in \eqref{eq:KNSigmaDelta}.
The inner and outer horizons are located at $r=r_-$ and $r=r_+$ respectively, with
\begin{align}
	r_\pm
	&=
		\frac12\left(
			r_s
			\pm \sqrt{r_s^2 - 4(a^2+r_Q^2)}
		\right)
	.
\end{align}
The surface $\Sigma_s$ intersects the outer horizon at some $v=v_0$ and $r=r_+$.
The angular speed on the surface is $\Omega_+ = \frac{a}{r_+^2+a^2}$, the surface gravity is $\kappa = \frac{r_+ - r_-}{2(r_+^2+a^2)}$, and the Hawking temperature is $T_H = \frac{\hbar\kappa}{2\pi}$.

The Kerr-Newman metric has zero scalar curvature, so the Einstein field equations read
\begin{align}
	R_{ab}
	&=
		8\pi G T_{ab}
	,
\end{align}
where $T_{ab}$ is the electromagnetic energy-momentum tensor \eqref{eq:EMT}.
The gauge field configuration solves the Maxwell equations \eqref{eq:emeom} outside the horizons, and is given by
\begin{align}
	A
	&=
		\frac{Q}{4\pi (\Sigma/r)}(-dv + a\sin^2\theta d\phi)
	,
\end{align}
with the field strength components,
\begin{align}
	F_{ab}
	&=
		\frac{Q}{4\pi \Sigma^2}
		\begin{pmatrix}
			0 & -(r^2-a^2\cos^2\theta) & 2ra^2\sin\theta\cos\theta & 0 \\
			\cdots & 0 & 0 & -(r^2-a^2\cos^2\theta) a\sin^2\theta \\
			\cdots & \cdots & 0 & 2ra(r^2+a^2)\sin\theta\cos\theta \\
			\cdots & \cdots & \cdots & 0
		\end{pmatrix}
	.
\end{align}
We can define the electromagnetic potential function in the exterior of the black hole to be
\begin{align}
	\Phi(r)
	&=
		\frac{Q r}{4\pi(r^2+a^2)}
	,
\end{align}
whose value on the horizon we refer to as $\Phi_+ = \Phi(r_+)$.

To find the entropy, we define $\xi^a$ as the Killing vector field normalized as \eqref{eq:parameterisation_xi}, and take the $U(1)$ gauge parameter to be proportional to
	the electromagnetic potential at the horizon as the following,
\begin{align}
	\xi^a\partial_a
	&=
		\frac{\hbar}{l_p T_H}\left(
			\frac{\partial}{\partial v}
			+ \Omega_+\frac{\partial}{\partial \phi}
		\right)
	,\qquad
	\epsilon
	=
		\frac{\hbar}{l_p T_H}\Phi_+
	.
	\label{eq:KNxiL}
\end{align}
We find that
\begin{align}
	\int_S -\iota_\xi\theta
	&=
		\frac{\hbar}{2l_p T_H}
		\Big[
			\delta M
			- \delta(\Phi(r)Q)|_{r=r_+}
		\Big]
		\label{eq:KNith}
	,\\
	\int_S -\hat F[\delta\xi,\delta\epsilon]
	&=
		\frac{\hbar}{l_p}
		\left[
			- \frac12(M+\Phi_+ Q)\delta\Big(\frac1{T_H}\Big)
			+ J\delta\Big(\frac{\Omega_+}{T_H}\Big)
			+ Q\delta\Big(\frac{\Phi_+}{T_H}\Big)
		\right]
		\label{eq:KNhFxL}
	.
\end{align}
A point worth noting is that $\delta \Phi(r)|_{r=r_+}$ is to be distinguished from $\delta \Phi_+$.
In the former expression the variation takes place before evaluation at $r=r_+$ and thus we take $\delta r=0$.
The latter expression is evaluated at $r=r_+$ before the variation, and thus we get contributions from $\delta r_+$ as $r_+$ is a function of $M$, $J$ and $Q$.
To see if the consistency condition \eqref{eq:consistency} is satisfied, we note the first law of Kerr-Newman thermodynamics is
\begin{align}
	\delta M
	&=
		T_H \delta S
		+ \Omega_+ \delta J
		+ \Phi_+ \delta Q
	,
\end{align}
and that the Smarr formula for Kerr-Newman black hole is
\begin{align}
	M
	&=
		2T_H S
		+ 2\Omega_+ J
		+ \Phi_+ Q
	.
\end{align}
These two formulae imply the following identity,
\begin{align}
	\frac{\delta M}{2T_H}
	- \frac{M}{2}\delta\Big(\frac{1}{T_H}\Big)
	+ J\delta\Big(\frac{\Omega_+}{T_H}\Big)
	- \frac{\Phi_+}{2T_H}\delta Q
	+ \frac{Q}{2}\delta\Big(\frac{\Phi_+}{T_H}\Big)
	&=
		0
	.
\end{align}
Applying this to the sum of \eqref{eq:KNith} and \eqref{eq:KNhFxL}, we obtain
\begin{align}
	\int_S -\iota_\xi\theta-\hat F[\delta\xi,\delta\epsilon]
	&=
		\int_S d\Omega
		\left.
		\delta \left[
			\frac{1}{4\pi r^2}
			\frac{\hbar Q}{2l_pT_H}\Big(
				\Phi_+
				- \Phi(r)
			\Big)
		\right]
		\right|_{r=r_+}
	,
\end{align}
which is indeed a total variation, so the consistency condition is satisfied.

The variation of the Noether charge conjugate to $\xi$ and $\epsilon$ can be computed using \eqref{eq:dQwCharge}.
We find that the integral of $\delta\hat F[\xi,\epsilon]$ on $S$ is
\begin{align}
	\int_S \delta \hat F[\xi,\epsilon]
	&=
		\int_S d\Omega
		\left.
		\delta\left[
		\frac{1}{4\pi r^2}
		\frac{\hbar}{l_p T_H}
		\left\{
			\frac{M}{2}
			- \Omega_+ J
			+ \left(\frac{\Phi(r)}{2} - \Phi_+\right) Q
		\right\}
		\right]
		\right|_{r=r_+}
	.
\end{align}
It follows from \eqref{eq:dQwCharge} that the variation of the charge is
\begin{align}
	\delta Q_{\xi,\epsilon}
	&=
		\delta\left[
		\frac{\hbar}{l_p T_H}\left(
			\frac{M}{2}
			- \Omega_+ J
			- \frac12\Phi_+ Q
		\right)
		\right]
	.
	\label{eq:KNdQ}
\end{align}
One can read off the full charge, and hence the entropy, as
\begin{align}
	S
	&=
		\frac{l_p}{\hbar} Q_{\xi,\epsilon}
	=
		\frac{1}{T_H}\left(
			\frac{M}{2}
			- \Omega_+ J
			- \frac12\Phi_+ Q
		\right)
	,
\end{align}
which is in accordance with the Smarr formula.

It is worth noting that the entropy coincides with the expression obtained by integrating $\hat F[\xi,\epsilon]$ directly on $S$ without taking a variation.
Also, one can compute $\delta Q_{\xi,\epsilon}$ using \eqref{eq:dQfromhG}, by taking $\xi^a$ and $\epsilon$ to be as in \eqref{eq:KNxiL},
	and $h_{ab}$ and $\delta A_a$ to be the variation in the metric and gauge field respectively coming from varying $M$, $J$ and $Q$.
The result is in agreement with \eqref{eq:KNdQ}, as anticipated.

\subsection{Inner horizon}

In this section, we test our formulae on the section $S'$ defined by a fixed $v=v_0$ and $r=r_-$, that is, the inner horizon of the Kerr-Newman black hole. 
For this to make sense, $S'$ has to be the boundary of a timelike slice $\Sigma'$, such that for instance $\int_{\Sigma'} \hat J = \int_{S'}\hat F$.
We will not attempt to address this issue here.
The results presented in this section should therefore be understood as formal extensions of our previous analysis,
	whose physical interpretation remains to be clarified.

The angular speed $\Omega_-$ on the inner horizon is $\Omega_- = \frac{a}{r_-^2+a^2}$.
The Killing vector $(k_-)^a\partial_a = \partial_v + \Omega_- \partial_\phi$ is null and geodesic on the inner horizon.
The geodesic equation is
\begin{align}
	(k_-)^b \nabla_b (k_-)^a = -\kappa_- (k_-)^a
	\label{eq:KN_rm_kgeo}
\end{align}
evaluated at $r=r_-$, with $\kappa_- = \frac{(r_+-r_-)}{2(r_-^2+a^2)}$ the surface gravity on the inner horizon.
Note the negative sign on the r.h.s.\ of \eqref{eq:KN_rm_kgeo} is arranged such that $\kappa_->0$.
We can define the temperature $T_-$ on the inner horizon to be
\begin{align}
	T_- = \frac{\hbar \kappa_-}{2\pi} = \frac{\hbar}{4\pi} \frac{(r_+-r_-)}{(r_-^2+a^2)}
	.
\end{align}
We also define the electric potential on the inner horizon as
\begin{align}
	\Phi_- = \Phi(r_-) = \frac{Q}{4\pi}\frac{r_-}{(r_-^2+a^2)}
	.
\end{align}
To make use of the entropy formula, we define the Killing vector $\xi_-$ and $U(1)$ gauge parameter $\epsilon_-$ to be
\begin{align}
	(\xi_-)^a\partial_a = \frac{\hbar}{l_p T_-} \left(\frac{\partial}{\partial v} + \Omega_- \frac{\partial}{\partial \phi}\right)
	,\qquad
	\epsilon_- = \frac{\hbar}{l_p T_-}\Phi_-
	.
\end{align}
Then we have the following integrals evaluated on $S'$,
\begin{align}
	\int_{S'} \delta\hat F[\xi_-,\epsilon_-]
	&=
		\left.
		\delta \left[
		\frac{\hbar}{l_pT_-}
		\left(
			\frac{M}{2}
			- \Omega_- J
			+ \left(
				\frac12\Phi(r)
				- \Phi_-
			\right)Q
		\right)
		\right]
		\right|_{r=r_-}
	,\\
	\int_{S'} -\iota_{\xi_-}\theta
	&=
		\frac{\hbar}{2l_p T_-}
		\Big[
			\delta M
			- \delta(\Phi(r)Q)|_{r=r_-}
		\Big]
	,\\
	\int_{S'} -\hat F[\delta\xi_-,\delta\epsilon_-]
	&=
		\frac{\hbar}{l_p}
		\left[
			- \frac12(M+\Phi_- Q)\delta\Big(\frac1{T_-}\Big)
			+ J\delta\Big(\frac{\Omega_-}{T_-}\Big)
			+ Q\delta\Big(\frac{\Phi_-}{T_-}\Big)
		\right]
	,
\end{align}
which are exactly analogous to the ones at $r=r_+$.
We can associate an entropy $S_-$ at the inner horizon with the area of the inner horizon.
The integral $\int_{S'} \hat G[k_-,\Phi_-]$ on $S'$ with the Killing vector $(k_-)^a\partial_a = \partial_v + \Omega_-\partial_\phi$ and the gauge parameter set to the potential $\Phi_-$
	yields the first law of thermodynamics at $r=r_-$,
\begin{align}
	\delta M
	&=
		T_-\delta S_-
		+ \Omega_- \delta J
		+ \Phi_- \delta Q
	.
\end{align}
This implies the inner horizon version of the Smarr formula
\begin{align}
	M
	&=
		2T_- S_-
		+ 2\Omega_- J
		+ \Phi_- Q
	.
\end{align}
It follows that
\begin{align}
	\frac{\delta M}{2T_-}
	- \frac{M}{2}\delta\Big(\frac{1}{T_-}\Big)
	+ J\delta\Big(\frac{\Omega_-}{T_-}\Big)
	- \frac{\Phi_-}{2T_-}\delta Q
	+ \frac{Q}{2}\delta\Big(\frac{\Phi_-}{T_-}\Big)
	&=
		0
	,
\end{align}
which can of course be checked explicitly.
This identity implies that
\begin{align}
	\int_{S'} -\iota_{\xi_-}\theta-\hat F[\delta\xi_-,\delta\epsilon_-]
	&=
		\left.
		\delta
		\left[
			\frac{\hbar Q}{2l_p T_-}(\Phi_--\Phi(r))
		\right]
		\right|_{r=r_-}
	.
\end{align}
Therefore, we have the infinitesimal charge
\begin{align}
	\delta Q_{\xi_-,\epsilon_-}
	&=
		\int_{S'}
		\delta\hat F[\xi_-,\epsilon_-]
		- \iota_{\xi_-}\theta
		- \hat F[\delta\xi_-,\delta\epsilon_-]
	\\ &=
		\frac{\hbar}{l_p}
		\delta\left(
			\frac {M}{2T_-}
			- \frac{\Omega_-J}{T_-}
			- \frac{\Phi_- Q}{2T_-}
		\right)
\end{align}
and the entropy associated with the inner horizon
\begin{align}
	S_-
	&=
		\frac{l_p}{\hbar}  Q_{\xi_-,\epsilon_-}
	=
		\frac {M}{2T_-}
		- \frac{\Omega_-J}{ T_-}
		- \frac{\Phi_- Q}{2T_-}
	,
\end{align}
in agreement with the Smarr formula.

\section{De Sitter Space}
\label{sec:DeSitter}

De Sitter spacetime is a vacuum solution of the Einstein equations with a positive cosmological constant.
It can be thought of globally as an $S^3$ that collapses from an infinite radius in the remote past, to a minimum
radius of $l=\sqrt{\frac{3}{\Lambda}}$ and then expands to infinite size in the remote future. The metric is
\begin{align}
	ds^2 = -d\tau^2 + l^2 \cosh^2\tfrac{\tau}{l}\bigl(d\chi^2 + \sin^2\chi(d\theta ^2 + \sin^2\theta d\phi^2) \bigr).
\end{align}
with $\chi,\theta$ and $\phi$ being the hyperspherical coordinates on $S^3$. Consider observers 
moving along timelike geodesics. The interior of their past light cone does not include the entirety of 
the spacetime, even in the limit as $\tau\rightarrow\infty$. In other words, these observers have a past event 
horizon. Furthermore, the event horizon of each observer is different. Around each observer's worldline,
one can construct a static metric that has a horizon in a way that is superficially similar to that 
encountered in the Schwarzschild solution. Suppose the observer is at the north pole of the $S^3$ where
$\chi=0$. The coordinate transformation into the static system is then
\begin{align} r &= l\cosh\frac{\tau}{l} \sin\chi\\
t &= \frac{l}{2}\ln\Biggl(\frac{\cosh\frac{\tau}{l}\cos\chi + \sinh\frac{\tau}{l}}{\cosh\frac{\tau}{l}\cos\chi - \sinh\frac{\tau}{l}}\Biggr)
\end{align}
resulting  in the metric
\begin{align}
	ds^2 = -\Bigl(1-\frac{r^2}{l^2}\Bigr)\, dt^2 + \Bigl(1-\frac{r^2}{l^2}\Bigr)^{-1}dr^2 + r^2(d\theta^2 +
	\sin^2\theta d\phi^2)
	, \label{eq:statdes}
\end{align}
where $0<r<l$.
The observer is now at $r=0$. There is nothing special about the choice of the north pole on $S^3$; by using the isometries of the spacetime we conclude that any geodesic observer would construct (\ref{eq:statdes})
around their worldlines.
The horizon is at $r=l$, and a routine calculation yields a temperature of 
$T=\frac{\hbar}{2\pi l}$. Gibbons and Hawking applied Euclidean field theory techniques and concluded that the
entropy of this horizon is $\frac{3\pi}{\hbar G\Lambda}$ \cite{Gibbons:1977mu}. The entropy is to be interpreted as a measure of 
the information behind each observer's horizon. It should be carefully noted that this entropy has nothing to
do with the constituents of gravitational collapse to form a black hole and its subsequent evaporation. It also has
nothing to do with the physics of spacetime singularities as de Sitter spacetime is devoid of singularities \cite{Hawking:1973uf}.
The entropy appears to be entirely due to the nature of spacetime. 
 
Now we can ask if our formula for the entropy of a region of space works for the de Sitter horizon. One 
can construct Eddington-Finkelstein coordinates to overcome the horizon coordinate singularity. Let $u = t - r^*$, where
\begin{align}
	r^* = \frac{l}{2}\ln\Biggl(\frac{l+r}{l-r}\Biggr).
\end{align}
The metric is now 
\begin{align}
	ds^2 = -\Bigl(1-\frac{r^2}{l^2}\Bigr)du^2 - 2du\,dr + r^2(d\theta^2+\sin^2\theta d\phi^2).
\end{align}
Just like the case of Schwarzschild black hole, $\frac{\partial}{\partial u}$ is a Killing vector that is null and 
geodesic on the horizon. With the normalization \eqref{eq:parameterisation_xi}, we find
\begin{align}
	\xi = \frac{2\pi l}{l_p} \frac{\partial}{\partial u}.
\end{align}
The entropy is then given by
\begin{align}
	S= \frac{l_p}{\hbar}Q_\xi = -\frac{1}{16\pi l_p}\int* d\xi
\end{align}
with the integral being taken over the horizon at some instant of retarded time $u$.
Evaluation yields
\begin{align}
	S=\frac{3\pi}{\hbar G \Lambda}
\end{align}
in agreement with the result of Gibbons and Hawking.

\section{Kottler Spacetime}\label{sec:SdS}

Next, we apply the formula to a case which has both a black hole horizon and a cosmological horizon:
	the Kottler spacetime \cite{Kottler:1918cxc}.

We consider a metric of the form
\begin{align}
	ds^2
	&=
		-V(r) dv^2
		+ 2dvdr
		+ r^2d\Omega_2^2
	,\qquad
	V(r)
	\equiv
	1-\frac{2GM}{r}-\frac{\Lambda}{3}r^2
	,
\end{align}
where $M$ is the mass of the black hole, $\Lambda>0$ is the positive cosmological constant, and we have defined the advanced time coordinate $v=t+r^*$ with the tortoise coordinate defined by $\frac{dr^*}{dr} = V^{-1}$.
The horizon radii are determined by the cubic equation
\begin{align}
	V(r)=1-\frac{2GM}{r}-\frac{\Lambda}{3}r^2 = 0
	.
	\label{eq:sdSV}
\end{align}
We restrict our attention to the case $0<\Lambda (3GM)^2<1$, for which the equation \eqref{eq:sdSV} has one negative solution $r_-$ and two positive solutions $r_1, r_2$ where $r_1<r_2$.
The spacetime exhibits a black hole horizon at $r=r_1$, a cosmological horizon at $r=r_2$, and a static region in between.
It is convenient to define a parameter $\beta$ by
\begin{align}
	\cos\beta
	&=
		\Lambda^{1/2}(3GM)
	,
	\qquad
	0<\beta<\frac\pi2
	,
\end{align}
in terms of which the three solutions are given by
\begin{align}
	r_-
	&=
		-\frac{2}{\sqrt\Lambda}\cos\frac\beta3
	,\qquad
	r_{1}
	=
		\frac{2}{\sqrt\Lambda}
		\cos\left(\frac\beta3+\frac\pi3\right)
	,\qquad
	r_{2}
	=
		\frac{2}{\sqrt\Lambda}
		\cos\left(\frac\beta3-\frac\pi3\right)
	.
\end{align}
Consider the time-like Killing vector $\partial_v$.
Its geodesic equation is
\begin{align}
	(\partial_v)^b\nabla_b (\partial_v)^a
	&=
		\frac{1}{2r}\left(1-\Lambda r^2\right)(\partial_v)^a
	=
		\pm\kappa (\partial_v)^a
	.
\end{align}
We evaluate this expression at the two horizon radii $r=r_i$, $i=1,2$, and refer to the quantities $\kappa_i$ as the surface gravities associated with the respective horizons,
	up to the fact that there is no natural way to normalize this null vector to unit length at ``infinity''.
By re-organizing terms in \eqref{eq:sdSV} to
\begin{align}
	\frac23\Lambda r^3- 2GM=-r(1-\Lambda r^2)
	,
	\label{eq:sdSV2}
\end{align}
one finds that $1-\Lambda r_1^2 > 0$ and $1-\Lambda r_2^2 < 0$.
Thus the expressions for $\kappa_i$ are
\begin{align}
	\kappa_1 = \frac{1}{2r_1}(1-\Lambda r_1^2)
	,\qquad
	\kappa_2 = \frac{1}{2r_2}(\Lambda r_2^2-1)
	.
\end{align}
We first compute the entropy associated with the black hole.
We define a Killing vector $\xi_1$ by rescaling $\partial_v$ using the normalization \eqref{eq:parameterisation_xi} at the black hole horizon $r=r_1$,
\begin{align}
	\xi_1^a\partial_a = \frac{2\pi}{l_p \kappa_1}\partial_v
	.
\end{align}
Then, by direct computation using \eqref{eq:sdSV2}, one obtains
\begin{align}
	(*d\xi)_{\theta\phi}
	&=
		\frac{2\pi}{l_p\kappa_1}
		\left(
			\frac23\Lambda r^3
			- 2GM
		\right)
	=
		\frac{2\pi}{l_p\kappa_1}
		r(\Lambda r^2-1)
	.
\end{align}
Therefore, we find that our formula for the entropy applied to any section $v=v_0$ of the black hole horizon $r=r_1$ evaluates to
\begin{align}
	S(r=r_1)
	&=
		-\frac{1}{16\pi l_p}\int_{r=r_1} *d\xi_1
	=
		\frac{4\pi r_1^2}{4 l_p^2}
	=
		\frac{A(r=r_1)}{4 G\hbar}
	,
\end{align}
which corresponds to the correct entropy from the area law.

To compute the entropy associated with the cosmological horizon $r=r_2$, we define another Killing vector $\xi_2$ normalized as \eqref{eq:parameterisation_xi} at $r=r_2$,
\begin{align}
	\xi_2^a\partial_a = -\frac{2\pi}{l_p \kappa_2}\partial_v
	.
\end{align}
The formula for the entropy evaluated at a section $v=v_0$ of the cosmological horizon $r=r_2$ evaluates to
\begin{align}
	S(r=r_2)
	&=
		-\frac{1}{16\pi l_p}\int_{r=r_2} *d\xi_2
	=
		\frac{4\pi r_2^2}{4 l_p^2}
	=
		\frac{A(r=r_2)}{4 G\hbar}
	,
\end{align}
which agrees with the area law associated with the cosmological horizon.

\section{Conclusions and Speculations}
\label{sec:Conc}

We have examined the proposal of \cite{Wald:1993nt,Iyer:1994ys} that the gravitational entropy of a horizon can be
described by a Noether charge. We believe we have put this proposal onto a more general footing by relating it to the
proposal of Bousso \cite{Bousso:2002ju} where the gravitational entropy of a spatial region is determined by the
lightsheet at its boundary. We have also examined the behavior of the variation of the Noether charge 
as described by covariant phase space methods. We find that our treatment reproduces the first law of black hole horizons
and we have illustrated this by explicit calculations in the Kerr-Newman spacetime. Our treatment also reproduces 
the entropy of the cosmological horizon in de Sitter space as first described by Gibbons and Hawking \cite{Gibbons:1977mu}.
Finally, we apply our method to black holes in de Sitter space and again find the expected entropy. 
In all cases we have here examined, the entropy is given by
\begin{align}
S= \frac{A}{4G\hbar}.
\end{align}
It should be noted that the entropy is of order $\hbar^{-1}$ and in general, one would expect modifications
to this formula from corrections of higher orders in $\hbar$.

The apparent generality of our proposal suggests two areas that require further exploration. The first is that our 
picture is sufficiently general that it appears that it could apply to more general regions of spacetime than horizons. 
In fact, it looks as if it could apply to any spatial region, as first suggested by Bousso. So we believe
that an examination of quantum extremal surfaces \cite{Engelhardt:2014gca} and cosmological models might
prove profitable. A second area to examine is, of course, contributions of higher order in $\hbar$. For example, 
a bath of thermal radiation has an entropy that comes from a one-loop contribution to the partition function
and so is of higher order in $\hbar$ than the contributions we have considered here. 
It should be noted, however, that these proposed directions are speculative, as the examples studied in this work exhibit stationary, highly symmetric horizons
	and therefore do not straightforwardly generalize to more complex settings.

All of this does not shed any light on the microscopic origin of gravitational entropy. It is often said 
that gravitational entropy represents the microstates of what is hidden behind a horizon. However, one could
reasonably expect that such a quantity would depend on the spectrum of elementary particles. To leading order
in $\hbar$, this does not happen although such a dependence will be found in higher order contributions.
Sometimes this difficulty is referred to as the species problem. 
To us, it seems more likely that gravitational entropy is telling us something fundamental about the
quantum nature of spacetime.

\acknowledgments
SC thanks Prahar Mitra, Richard Myers and Antony Speranza for discussions.
MP would like to thank the STFC for financial support under grant ST/L000415/1.
SC is supported by the European Research Council (ERC) under the European Union’s Horizon 2020 research and innovation programme (grant agreement No 852386).

\appendix
\section{Appendix --- Interlude to the variation of $\xi$.}\label{appendix}

In this section, we briefly review the black hole entropy computation in the literature and introduce the role of the variation of the vector field in this context.

For rotating black holes, the vector field $\xi^a$ takes the form
\begin{align}
	\xi^a = {\bf t}^a + \Omega_+ {\boldsymbol \phi}^a
\end{align}
where ${\bf t} = \partial_t = \partial_v$ and ${\boldsymbol \phi} = \partial_\phi$ are the two Killing vectors.
The diffeomorphism $\xi$ is a symmetry in the sense that ${\cal L}_\xi \phi=0$ on all fields, so
\begin{align}
	\omega(\phi,\delta\phi,{\cal L}_\xi\phi) = 0
	.
\end{align}
Consider a Cauchy slice $\Sigma$ bounded by two spheres, one at $r=\infty$ and the other at the bifurcation surface $B$, where $v=-\infty$ and $r=r_+$.
Integrating $\omega(\phi,\delta\phi,{\cal L}_\xi\phi)$ on $\Sigma$ yields zero, so
\begin{align}
	0
	&=
		\delta Q_\xi
	=
		\left(-\int_B+\int_\infty\right ) (\delta \hat F[\xi]-\iota_\xi\theta - \hat F[\delta\xi])
	,
\end{align}
where the minus sign accounts for the opposite orientation.
Thus
\begin{align}
	\int_B \delta \hat F[\xi]-\iota_\xi\theta - \hat F[\delta\xi]
	=
	\int_\infty \delta \hat F[\xi]-\iota_\xi\theta - \hat F[\delta\xi]
	.
\end{align}
The last term is missing in the earlier literature \cite{Lee:1990nz,Wald:1993nt,Iyer:1994ys,Wald:1999wa} as they consider vectors such that $\delta\xi^a=0$.
The authors compute the integral at $\infty$ to get $\delta M-\Omega_+\delta J$, and use this to deduce that the integral on $B$ is $T\delta S$.
On the bifurcation surface, $\iota_\xi\theta$ does not contribute to the integral.
This is because
\begin{enumerate}
	\item by linearity $\iota_\xi\theta = \iota_{\bf t}\theta + \Omega_+ \iota_{\boldsymbol\phi} \theta$.
	\item ${\bf t}=\partial_v$ vanishes on $B$, so $\iota_{\bf t}\theta=0$; for instance if we define $\tilde v\equiv e^{v}$, then $\partial_v = \tilde v \partial_{\tilde v}$ and this vanishes at $B$ where $\tilde v=0$.
	\item ${\boldsymbol\phi}$ is tangent to $B$, so the pullback of $\iota_{\boldsymbol \phi}\theta$ to $B$ is zero.
\end{enumerate}
It follows that $\int_B \delta \hat F[\xi]$ is $T\delta S$.
Thus, up to the Hawking temperature, $\int_B \hat F[\xi]$ without the $\delta$ is the full entropy.
If we consider the integral $\int_S \hat F[\xi]$ on any section $S$ at $v=v_0$, $r=r_+$, then using $\hat J = d\hat F$ and Stokes' theorem we can write
\begin{align}
	\int_S \hat F[\xi] - \int_B \hat F[\xi] = \int_{\overline{BS}} \hat J[\xi]
\end{align}
where $\overline{BS}$ is the lower segment of the outer horizon extending from $B$ ($v=-\infty$) to $S$ ($v=v_0$).
But the r.h.s.\ is zero, since by definition
\begin{align}
	\hat J[\xi] = \theta(\phi,{\cal L}_\xi\phi) - \iota_\xi L
	,
\end{align}
where ${\cal L}_\xi\phi=0$ implies $\theta(\phi,{\cal L}_\xi\phi)=0$ by linearity, and the pullback of $\iota_\xi L=0$ to $\overline{BS}$ vanishes since $\xi$ is tangent to the future horizon.
Therefore
\begin{align}
	\int_S \hat F[\xi] = \int_B \hat F[\xi]
	,
\end{align}
and that the entropy, up to Hawking temperature, can be computed using any section $v=v_0$ of the horizon.

Note that this method does not work if we try to integrate expressions like the following on $S$ (not the bifurcation surface $B$):
\begin{align}
	\int_S \delta \hat F[\xi] - \iota_\xi\theta
	.
\end{align}
This expression does not give us the entropy, because $\iota_\xi\theta$ does not necessarily vanish on $S\neq B$.
What we find is that the following expression
\begin{align}
	\int_S \delta \hat F[\xi] - \iota_\xi\theta - \hat F[\delta\xi]
\end{align}
yields (the variation of) the correct entropy.
From the previous argument that the entropy is given by $\int_S \hat F$, this implies that the last two terms on the r.h.s.\ cancel out.
This is exactly what we find for Schwarzschild, Kerr, and Kerr-Newman cases.

\vskip 4cm

\bibliographystyle{jhep}
\bibliography{references}

\providecommand{\href}[2]{#2}\begingroup\raggedright\begin{thebibliography}{10}

\bibitem{Bardeen:1973gs}
J.M.~Bardeen, B.~Carter and S.W.~Hawking, \emph{{The Four laws of black hole
  mechanics}}, \href{https://doi.org/10.1007/BF01645742}{\emph{Commun. Math.
  Phys.} {\bfseries 31} (1973) 161}.

\bibitem{Bekenstein:1973ur}
J.D.~Bekenstein, \emph{{Black holes and entropy}},
  \href{https://doi.org/10.1103/PhysRevD.7.2333}{\emph{Phys. Rev. D} {\bfseries
  7} (1973) 2333}.

\bibitem{Hawking:1975vcx}
S.W.~Hawking, \emph{{Particle Creation by Black Holes}},
  \href{https://doi.org/10.1007/BF02345020}{\emph{Commun. Math. Phys.}
  {\bfseries 43} (1975) 199}.

\bibitem{Wald:1993nt}
R.M.~Wald, \emph{{Black hole entropy is the Noether charge}},
  \href{https://doi.org/10.1103/PhysRevD.48.R3427}{\emph{Phys. Rev. D}
  {\bfseries 48} (1993) R3427}
  [\href{https://arxiv.org/abs/gr-qc/9307038}{{\ttfamily gr-qc/9307038}}].

\bibitem{Jacobson:1993vj}
T.~Jacobson, G.~Kang and R.C.~Myers, \emph{{On black hole entropy}},
  \href{https://doi.org/10.1103/PhysRevD.49.6587}{\emph{Phys. Rev. D}
  {\bfseries 49} (1994) 6587}
  [\href{https://arxiv.org/abs/gr-qc/9312023}{{\ttfamily gr-qc/9312023}}].

\bibitem{Iyer:1994ys}
V.~Iyer and R.M.~Wald, \emph{{Some properties of Noether charge and a proposal
  for dynamical black hole entropy}},
  \href{https://doi.org/10.1103/PhysRevD.50.846}{\emph{Phys. Rev. D} {\bfseries
  50} (1994) 846} [\href{https://arxiv.org/abs/gr-qc/9403028}{{\ttfamily
  gr-qc/9403028}}].

\bibitem{Dong:2013qoa}
X.~Dong, \emph{{Holographic Entanglement Entropy for General Higher Derivative
  Gravity}}, \href{https://doi.org/10.1007/JHEP01(2014)044}{\emph{JHEP}
  {\bfseries 01} (2014) 044} [\href{https://arxiv.org/abs/1310.5713}{{\ttfamily
  1310.5713}}].

\bibitem{Wall:2015raa}
A.C.~Wall, \emph{{A Second Law for Higher Curvature Gravity}},
  \href{https://doi.org/10.1142/S0218271815440149}{\emph{Int. J. Mod. Phys. D}
  {\bfseries 24} (2015) 1544014}
  [\href{https://arxiv.org/abs/1504.08040}{{\ttfamily 1504.08040}}].

\bibitem{Mishra:2017sqs}
A.~Mishra, S.~Chakraborty, A.~Ghosh and S.~Sarkar, \emph{{On the physical
  process first law for dynamical black holes}},
  \href{https://doi.org/10.1007/JHEP09(2018)034}{\emph{JHEP} {\bfseries 09}
  (2018) 034} [\href{https://arxiv.org/abs/1709.08925}{{\ttfamily
  1709.08925}}].

\bibitem{Hollands:2024vbe}
S.~Hollands, R.M.~Wald and V.G.~Zhang, \emph{{Entropy of dynamical black
  holes}}, \href{https://doi.org/10.1103/PhysRevD.110.024070}{\emph{Phys. Rev.
  D} {\bfseries 110} (2024) 024070}
  [\href{https://arxiv.org/abs/2402.00818}{{\ttfamily 2402.00818}}].

\bibitem{Gao:2001ut}
S.~Gao and R.M.~Wald, \emph{{The 'Physical process' version of the first law
  and the generalized second law for charged and rotating black holes}},
  \href{https://doi.org/10.1103/PhysRevD.64.084020}{\emph{Phys. Rev. D}
  {\bfseries 64} (2001) 084020}
  [\href{https://arxiv.org/abs/gr-qc/0106071}{{\ttfamily gr-qc/0106071}}].

\bibitem{Ashtekar:2024stm}
A.~Ashtekar and S.~Speziale, \emph{{Null infinity and horizons: A new approach
  to fluxes and charges}},
  \href{https://doi.org/10.1103/PhysRevD.110.044049}{\emph{Phys. Rev. D}
  {\bfseries 110} (2024) 044049}
  [\href{https://arxiv.org/abs/2407.03254}{{\ttfamily 2407.03254}}].

\bibitem{Gao:2003ys}
S.~Gao, \emph{{The First law of black hole mechanics in Einstein-Maxwell and
  Einstein-Yang-Mills theories}},
  \href{https://doi.org/10.1103/PhysRevD.68.044016}{\emph{Phys. Rev. D}
  {\bfseries 68} (2003) 044016}
  [\href{https://arxiv.org/abs/gr-qc/0304094}{{\ttfamily gr-qc/0304094}}].

\bibitem{Frodden:2017qwh}
E.~Frodden and D.~Hidalgo, \emph{{Surface Charges for Gravity and
  Electromagnetism in the First Order Formalism}},
  \href{https://doi.org/10.1088/1361-6382/aa9ba5}{\emph{Class. Quant. Grav.}
  {\bfseries 35} (2018) 035002}
  [\href{https://arxiv.org/abs/1703.10120}{{\ttfamily 1703.10120}}].

\bibitem{Adami:2021nnf}
H.~Adami, D.~Grumiller, M.M.~Sheikh-Jabbari, V.~Taghiloo, H.~Yavartanoo and
  C.~Zwikel, \emph{{Null boundary phase space: slicings, news {\&} memory}},
  \href{https://doi.org/10.1007/JHEP11(2021)155}{\emph{JHEP} {\bfseries 11}
  (2021) 155} [\href{https://arxiv.org/abs/2110.04218}{{\ttfamily
  2110.04218}}].

\bibitem{Chandrasekaran:2021vyu}
V.~Chandrasekaran, E.E.~Flanagan, I.~Shehzad and A.J.~Speranza, \emph{{A
  general framework for gravitational charges and holographic
  renormalization}},
  \href{https://doi.org/10.1142/S0217751X22501056}{\emph{Int. J. Mod. Phys. A}
  {\bfseries 37} (2022) 2250105}
  [\href{https://arxiv.org/abs/2111.11974}{{\ttfamily 2111.11974}}].

\bibitem{Barnich:2001jy}
G.~Barnich and F.~Brandt, \emph{{Covariant theory of asymptotic symmetries,
  conservation laws and central charges}},
  \href{https://doi.org/10.1016/S0550-3213(02)00251-1}{\emph{Nucl. Phys. B}
  {\bfseries 633} (2002) 3}
  [\href{https://arxiv.org/abs/hep-th/0111246}{{\ttfamily hep-th/0111246}}].

\bibitem{Barnich:2011mi}
G.~Barnich and C.~Troessaert, \emph{{BMS charge algebra}},
  \href{https://doi.org/10.1007/JHEP12(2011)105}{\emph{JHEP} {\bfseries 12}
  (2011) 105} [\href{https://arxiv.org/abs/1106.0213}{{\ttfamily 1106.0213}}].

\bibitem{Bousso:2002ju}
R.~Bousso, \emph{{The Holographic principle}},
  \href{https://doi.org/10.1103/RevModPhys.74.825}{\emph{Rev. Mod. Phys.}
  {\bfseries 74} (2002) 825}
  [\href{https://arxiv.org/abs/hep-th/0203101}{{\ttfamily hep-th/0203101}}].

\bibitem{Gibbons:1977mu}
G.W.~Gibbons and S.W.~Hawking, \emph{{Cosmological Event Horizons,
  Thermodynamics, and Particle Creation}},
  \href{https://doi.org/10.1103/PhysRevD.15.2738}{\emph{Phys. Rev. D}
  {\bfseries 15} (1977) 2738}.

\bibitem{Wald:1999wa}
R.M.~Wald and A.~Zoupas, \emph{{A General definition of 'conserved quantities'
  in general relativity and other theories of gravity}},
  \href{https://doi.org/10.1103/PhysRevD.61.084027}{\emph{Phys. Rev. D}
  {\bfseries 61} (2000) 084027}
  [\href{https://arxiv.org/abs/gr-qc/9911095}{{\ttfamily gr-qc/9911095}}].

\bibitem{York:1972sj}
J.W.~York, Jr., \emph{{Role of conformal three geometry in the dynamics of
  gravitation}}, \href{https://doi.org/10.1103/PhysRevLett.28.1082}{\emph{Phys.
  Rev. Lett.} {\bfseries 28} (1972) 1082}.

\bibitem{Gibbons:1976ue}
G.W.~Gibbons and S.W.~Hawking, \emph{{Action Integrals and Partition Functions
  in Quantum Gravity}},
  \href{https://doi.org/10.1103/PhysRevD.15.2752}{\emph{Phys. Rev. D}
  {\bfseries 15} (1977) 2752}.

\bibitem{Smarr:1972kt}
L.~Smarr, \emph{{Mass formula for Kerr black holes}},
  \href{https://doi.org/10.1103/PhysRevLett.30.71}{\emph{Phys. Rev. Lett.}
  {\bfseries 30} (1973) 71}.

\bibitem{Hajian:2015xlp}
K.~Hajian and M.M.~Sheikh-Jabbari, \emph{{Solution Phase Space and Conserved
  Charges: A General Formulation for Charges Associated with Exact
  Symmetries}}, \href{https://doi.org/10.1103/PhysRevD.93.044074}{\emph{Phys.
  Rev. D} {\bfseries 93} (2016) 044074}
  [\href{https://arxiv.org/abs/1512.05584}{{\ttfamily 1512.05584}}].

\bibitem{Hawking:1973uf}
S.W.~Hawking and G.F.R.~Ellis, \emph{{The Large Scale Structure of
  Space-Time}}, Cambridge Monographs on Mathematical Physics, Cambridge
  University Press (2, 2023),
  \href{https://doi.org/10.1017/9781009253161}{10.1017/9781009253161}.

\bibitem{Kottler:1918cxc}
F.~Kottler, \emph{{{\"U}ber die physikalischen Grundlagen der Einsteinschen
  Gravitationstheorie}},
  \href{https://doi.org/10.1002/andp.19183611402}{\emph{Annalen Phys.}
  {\bfseries 361} (1918) 401}.

\bibitem{Engelhardt:2014gca}
N.~Engelhardt and A.C.~Wall, \emph{{Quantum Extremal Surfaces: Holographic
  Entanglement Entropy beyond the Classical Regime}},
  \href{https://doi.org/10.1007/JHEP01(2015)073}{\emph{JHEP} {\bfseries 01}
  (2015) 073} [\href{https://arxiv.org/abs/1408.3203}{{\ttfamily 1408.3203}}].

\bibitem{Lee:1990nz}
J.~Lee and R.M.~Wald, \emph{{Local symmetries and constraints}},
  \href{https://doi.org/10.1063/1.528801}{\emph{J. Math. Phys.} {\bfseries 31}
  (1990) 725}.

\end{thebibliography}\endgroup

\end{document}